\documentclass[twocolumn,preprintnumbers,nofootinbib,prl,showpacs]{revtex4}

\usepackage{graphicx}
\usepackage{graphics}
\usepackage{dsfont}
\usepackage{epsfig}
\usepackage{amsmath,amssymb,amsthm,amscd}

\usepackage{hyperref}
\newcommand{\beq}{\begin{equation}}
\newcommand{\eeq}{\end{equation}}
\newcommand{\be}{B_\oplus}

\def\be{\begin{equation}}
\def\ee{\end{equation}}
\def\baray{\begin{eqnarray}}
\def\earay{\end{eqnarray}}
\def\bes{\begin{eqnarray}}
\def\ens{\end{eqnarray}}
\def\ba{\begin{eqnarray}}
\def\ea{\end{eqnarray}}

\def\bes {\begin{eqnarray}}
\def \ens {\end{eqnarray}}

\begin{document}

\newcommand{\bea}{\begin{eqnarray}}
\newcommand{\eea}{\end{eqnarray}}
\newcommand{\barr}{\begin{array}}
\newcommand{\earr}{\end{array}}

\pagestyle{plain}

\preprint{USTC-ICTS-15-05}

\title{New Exact Quantization Condition for Toric Calabi-Yau Geometries}

\author{Xin Wang\footnote{email: wxin@mail.ustc.edu.cn}, 
Guojun Zhang\footnote{email: zgj1@mail.ustc.edu.cn}, 
Min-xin Huang\footnote{email: minxin@ustc.edu.cn} 
}

\affiliation{Interdisciplinary Center for Theoretical Study, School of Physical Sciences \\     University of Science and Technology of China,  Hefei, Anhui 230026, China
}


\begin{abstract}
We propose a new exact quantization condition for a class of quantum mechanical systems derived from local toric Calabi-Yau three-folds. Our proposal includes all contributions to the energy spectrum which are non-perturbative in the Planck constant, and is much simpler than the available quantization condition in the literature. We check that our proposal is consistent with previous works and implies non-trivial relations among the topological Gopakumar-Vafa invariants of the toric Calabi-Yau geometries. Together with the recent developments, our proposal opens a new avenue in the long investigations at the interface of geometry, topology and quantum mechanics.     
\end{abstract}
\maketitle

Non-perturbative phenomena are ubiquitous in quantum theories.  A classic example is the quantum mechanics of a non-relativistic particle moving in a double-well potential. Here the non-perturbative effects come from instantons tunneling between different local minima of the potential or its analytic continuation, and are proportional to positive powers of $\exp (-\frac{1}{\hbar})$, see e.g.  chapter 7 of the book \cite{Coleman}. It is not easy to find exact quantization conditions which include all instanton contributions. In a series of seminar works, Zinn-Justin et al proposed such exact quantization conditions for certain quantum mechanical systems with double-well and periodic potentials \cite{ZinnJustin:1981, ZinnJustin:2004a}. The eventual proof of the quantization conditions has produced tremendous advances in modern mathematical physics, entailing the developments in the advanced theory of resurgence \cite{Ecalle, Pham}.  

It has long been also a fascinating idea to connect the study of geometries with quantum mechanics, which in a sense lead to the so called quantum geometry, see e.g. \cite{ADKMV, Aganagic:2011}. Here we shall focus on the Calabi-Yau geometries, which have been the main playground for topological string theory. This is a big subject and we shall not go into the details in this Letter. The main ingredient that we will need is a type of topological invariants of such geometries, known as the refined Gopakumar-Vafa invariants, which have been computed systematically in many cases.   

Recently, there have some considerable interests in the studies of a novel type of quantum mechanical systems, where the Hamiltonian is derived from toric Calabi-Yau geometries, and consists of exponentials of linear combinations of the quantum position and momentum operators. For example, the simplest non-trivial example is derived from local $\mathbb{P}^2$ Calabi-Yau geometry, with the following exponential of the Hamiltonian 
\begin{eqnarray} \label{localP2}
e^{\hat{H}} = e^{\hat{x}} + e^{\hat{p}} + e^{-\hat{x}-\hat{p}} ,
\end{eqnarray} 
where the $\hat{x}, \hat{p}$ operators satisfy the canonical commutation relation $[\hat{x},\hat{p}]=i\hbar$. The Hamiltonian is Hermitian and bounded below by the classical minimum $H\geq \log(3)$, so we expect it has well-defined energy spectrum for any Planck constant $\hbar\geq 0$. Based on earlier works on perturbative results \cite{Aganagic:2011}, and the ABJM (Aharony-Bergman-Jafferis-Maldacena) matrix model \cite{Hatsuda:2012, Hatsuda:2013, Kallen:2013}, one of the co-author of the present Letter was able to establish the relation between the exact discrete energy spectrum of quantum systems of general toric Calabi-Yau geometries and the corresponding topological Gopakumar-Vafa invariants \cite{Huang:2014}. The formulas for some low order non-perturbative contributions to the exact quantization condition are determined by numerical analysis. Soon afterward, a much more powerful conjecture for the spectral determinant of such quantum system was proposed by Grassi, Hatsuda and Mari\~{n}o \cite{Grassi:2014}, which we shall call the GHM conjecture. Our formulas in \cite{Huang:2014} can then be analytically derived from the GHM conjecture, and served as its non-trivial checks. For some more recent developments, see e.g. \cite{Moriyama:2014, Kashaev:2015, Marino:2015, Hatsuda:2015, Kashaev:2015a}.  

Our results in this Letter may have broad interdisciplinary impacts on many research fields. Firstly, we know that exact results in quantum theories are often very useful for analyzing strong coupling phenomena. However, without supersymmetry, such exact results are very rare to come by. It is thus quite both astonishing and satisfying for us to see that a seemingly simple non-supersymmetric quantum mechanical system such as  (\ref{localP2}) could have a deep connection to the Calabi-Yau geometry, and the exact quantization condition can be written in terms of topological invariants of the corresponding geometry.  Furthermore, due to the relation with topological strings, our proposal may provide useful insights into non-perturbative formulations of string/M theory, which still is the main challenge of the field. Finally, although we don't have concrete examples at the moment, these quantum systems may well be related to some condense matter systems as many exactly solvable systems do. If this is the case, it is not unthinkable that our proposal could help to understand the corresponding experimental results.

Now we shall present the main result of the Letter. We will use the local $\mathbb{P}^{2}$ model  in (\ref{localP2}) as the main illustrative example.  We conjecture that the discrete energy spectrum satisfies the following quantization condition 
\begin{eqnarray}  \label{BScondition}
\textrm{vol}_{\rm p}(E, \hbar) + \textrm{vol}_{\rm np}(E, \hbar) = 2\pi \hbar(n+\frac{1}{2}), 
\end{eqnarray}
where $n=0,1,2, \cdots$ is the discrete energy level, and the formulas for perturbative and non-perturbative parts 
$\textrm{vol}_{\rm p}(E, \hbar)$ and $\textrm{vol}_{\rm np}(E, \hbar)$ are 
\begin{eqnarray} \label{vol}
\textrm{vol}_{\rm p}(E,\hbar ) &=& \frac{\tilde{t}^2-\pi^2}{2} -\frac{\hbar^2}{8}+\hbar   f_{NS} (\tilde{t} , \hbar), 
 \nonumber \\
\textrm{vol}_{\rm np}(E,\hbar ) &=& \hbar   f_{NS} (\frac{2\pi \tilde{t}}{\hbar} , \frac{4\pi^2}{\hbar}). 
\end{eqnarray}
Some more explanation of the notations is necessary. Here $\tilde{t}$ is known as the quantum A-period of the Calabi-Yau geometry, and can be calculated straightforwardly from the Hamiltonian \cite{Aganagic:2011}. The Schr\"{o}dinger equation with a wave function $\psi(x)$ can be transformed as
\begin{eqnarray}
\frac{zV(Xq)}{Xq^{\frac{1}{2}}} +X-1+\frac{1}{V(X)}=0 , 
\end{eqnarray}
where the parameters are $z=e^{-3E}, X=e^x, q=e^{i\hbar}$ and $V(X)=\frac{\psi(x)}{\psi(x-i\hbar)}$. The function $V(X)$ can be solved perturbatively in small $z$ expansion, and the the quantum A-period is computed by its residue around $X=0$ as
\begin{eqnarray} \label{deformedA}
\tilde{t} (E,\hbar) &=& \log(z) + 3\oint \frac{d X}{2\pi i X} \log(V(X))    \nonumber \\ 
&=&  -3E+ 3 (e^{\frac{i\hbar}{2}} + e^{-\frac{i\hbar}{2}}) e^{-3E} 
+ 3[  e^{2i\hbar} + e^{- 2i\hbar}
\nonumber \\ && + \frac{7}{2} (e^{i\hbar} + e^{- i\hbar})  + 6  ] e^{-6E} +\mathcal{O}(e^{-9E}) . 
\end{eqnarray}
It is convenient to introduce an index parameter $r$ for toric Calabi-Yau geometries, such that the quantum A-period is $\tilde{t} (E,\hbar)  \sim  - rE$ for large $E$ and can be expressed as a power series of $e^{-rE}$. For the local $\mathbb{P}^2$ model we have $r= 3$.  The function $f_{NS} (\tilde{t} , \hbar)$ is proportional to the derivative of world-sheet instanton part of the topological string amplitude in the Nekrasov-Shatashvili limit with respect to the quantum A-period $\tilde{t}$, and can be expressed in terms of the refined Gopakumar-Vafa $n^{d}_{j_L,j_R}$ as
\begin{eqnarray} \label{fNS}
f_{NS} (\tilde{t} , \hbar) &=& - \frac{r}{2} \sum_{j_L,j_R} \sum_{w,d=1}^{\infty}   (-1)^{2j_L+2j_R+rwd}   n^{d}_{j_L,j_R}    e^{wd\tilde{t}}   \nonumber \\ & \times & \frac{d}{w} \cdot
\frac{\sin \frac{w\hbar(2j_R+1)}{2} \sin \frac{w\hbar(2j_L+1)}{2}}{ \sin^3 (\frac{w\hbar}{2})}. 
\end{eqnarray}
Here the indices $j_L,j_R$ are non-negative (half-)integers corresponding the representation under the decomposition of 5D little group $SO(4)\cong SU(2)_L \times SU(2)_R$, and the degree $d$ denotes the element of second homology class of the Calabi-Yau geometry. The refined Gopakumar-Vafa invariants $n^{d}_{j_L,j_R}$ are non-negative integers and only non-vanishing up to finite $j_L, j_R$ for a given degree $d$. They are first computed by the method of refined topological vertex \cite{IKV}. The data are also available in e.g. our previous paper \cite{Huang:2014}.

The power series in (\ref{deformedA}), (\ref{fNS}) are convergent when the energy $E\geq \log(3)$, i.e. no less than the classical minimum. Therefore for any energy level $n$ and Planck constant $\hbar$, we can solve for the corresponding energy $E$ from our proposal of the quantization relation (\ref{BScondition}). This can be done  in principle to arbitrary numerical precision. In the spirit of Bohr-Sommerfeld quantization, the left hand side of our proposal (\ref{BScondition}) can then be regarded as the exact non-perturbative definition of quantum phase volume. In the classical limit $\hbar=0$, we have $\textrm{vol}_{\rm np}(E, 0)=0$ and the perturbative part is simply the bounded area $\textrm{vol}_{\rm p}(E, 0)= \int_{H(x,p)\leq E} dxdp$, consistent with the usual convention \cite{Huang:2014}.     

The use of Nekrasov-Shatashvili limit for the perturbative part $\textrm{vol}_{\rm p}(E,\hbar)$ is already proposed in \cite{Aganagic:2011}, and also in earlier works on Seiberg-Witten theory \cite{NS, MM}. The various constants are determined in our previous paper \cite{Huang:2014}. Our new contribution in this Letter is that the non-perturbative contributions can be simply obtained by the replacement of variables 
\begin{equation} \label{transform}
\tilde{t}\rightarrow \frac{2\pi \tilde{t}}{\hbar}, ~~~~ \hbar\rightarrow \frac{4\pi^2}{\hbar},
\end{equation}
in the instanton contribution $f_{NS} (\tilde{t} , \hbar)$. Our conjecture is inspired by earlier works on non-perturbative completion of topological string theory \cite{Pasquetti:2009, Lockhart:2012, Couso-Santamaria:2014}, but the specific contexts and details are somewhat different. Here our quantum mechanical system is already well-defined non-perturbatively for any Planck constant $\hbar$, and as such our proposal can be tested precisely by numerical computations. We can choose any complete orthonormal basis of quantum wave functions, such as the eigenfunctions of a harmonic oscillator, truncate to a finite energy level, and diagonalize the matrix for $e^{\hat{H}}$ to find the energy eigenvalues. It is found empirically that as we increase the matrix size, the energy eigenvalues do converge  \cite{Huang:2014}. This is justified later in  \cite{Kashaev:2015}, which proves that the $\exp (-\hat{H})$ is indeed a well-defined trace class operator. So in principle we can also compute the energy eigenvalues to any numerical precision in this way, and use the results to test the conjectural quantization conditions. We shall check below that the proposal for non-perturbative phase volume agrees with the numerically fitted formulas obtained up to the 5th order in our previous paper \cite{Huang:2014}.

It is easy to see that the perturbative phase volume has singularities when $\hbar/\pi$ is a rational number. As a consistency check, since the quantum spectrum is well-defined for any $\hbar$, these singularities must be cancelled by the non-perturbative contributions. Motivated by studies from the ABJM matrix model, Kallen and Mari\~{n}o proposed to use the conventional topological string amplitude without refinement, which is also known as the unrefined amplitude and is a different limit from the Nekrasov-Shatashvili limit, to cancel the singularities \cite{Kallen:2013}. Here we can similarly check that our new proposal for the non-perturbative contributions to the phase volume can also cancel the singularities, and we also need to use the property that for the a local toric Calabi-Yau geometry, the non-vanishing  refined Gopakumar-Vafa invariants $n^{d}_{j_L,j_R}$ always has $2j_L+2j_R+rd$ as an odd integer, with $r=3$ for the local $\mathbb{P}^2$ model. We emphasize that the singularity cancellations from the two methods are independent, but our proposal has the advantage that it also captures the non-singular contributions.  

We can check that our new conjecture is consistent with the previous GHM conjecture in \cite{Grassi:2014}. A quantization condition, which can be also written in the Bohr-Sommerfeld form, has been derived there, by setting the spectral determinant to zero. The perturbative part is identical to the one in (\ref{vol}), but the non-perturbative part is more complicated, consisting of the unrefined amplitude and an infinite series of corrections. The unrefined amplitude is responsible for singularity cancellations, while the infinite series of corrections are non-singular and have to be calculated order by order from the spectral determinant formula. To write down the formula for the local $\mathbb{P}^2$ model, we shall follow \cite{Grassi:2014} and defined some functions as
\begin{eqnarray} \label{formula7}
 f &:= &  \frac{ n^{d}_{j_L,j_R}}{4w} (-1)^{2j_L+2j_R+rwd} Q^{wd}   \nonumber \\ &\times&
  \frac{(2j_R+1) \sin [4x w (2j_L+1) ]}{\sin^2 (2x w)\sin(4x w ) },
 \nonumber \\ 
 f_c (n) & := &   \sum_{j_L,j_R} \sum_{w,d=1}^{\infty} [ \cos (2r(2n + 1) x wd ) 
  \nonumber \\  &&  - \cos(2r x wd) ]\cdot f , 
  \nonumber \\ 
   f_s (n) & := & \sum_{j_L,j_R} \sum_{w,d=1}^{\infty}   [ \sin (2r (2n + 1) x wd )
   \nonumber \\  && - (2n+1) \sin (2rx wd) ] \cdot f, 
\end{eqnarray}
where we denote $x :=  \frac{\pi^2}{\hbar} $ and $Q:= e^{\frac{2 \pi \tilde{t} }{\hbar} }$. The GHM quantization condition gives the non-perturbative phase volume 
\begin{eqnarray} \label{GHMvol}
 \textrm{vol}_{\rm np}^{\rm GHM}(E,\hbar) = - 2 \hbar  \sum_{j_L,j_R} \sum_{w,d=1}^{\infty}  f\cdot \sin( 2r x wd )    +\lambda    , ~~~~~  \end{eqnarray}
where $\lambda := \sum_{k=1}^{\infty} c_k(x) Q^k $  is the correction term and is determined by the following equation 
\begin{eqnarray} \label{formula9}
&& \sum_{n=0}^{\infty}  \sin [3n(n + 1)(2n + 1)x + f_s(n) + (n + \frac{1}{2})\lambda ]  \nonumber \\ &\times & Q^{\frac{3n(n + 1)}{2}} (-1)^n e^{ f_c(n)} =0 
\end{eqnarray} 
The correction terms can be solved perturbatively by expansion of the equation in powers of $Q$.  Using the data of refined Gopakumar-Vafa invariants, the solutions for the first few coefficients are  
 \begin{eqnarray}  \label{coefficients}
   && c_1(x) = c_2(x) = 0, ~~~ c_3(x) = 2 \sin(18 x),  \nonumber \\ && c_4(x) = \frac{6\sin^2(6x)\sin(24 x)}{\sin^2(2x)},
  \nonumber \\ && c_5(x) =    \frac{6\sin(6x)\sin(30 x)}{\sin^2(2x)} [16\sin(2x)\sin^2(6x)  \nonumber \\ && + 
   20\sin(2x)\sin(10x)\sin(6x) + 7\sin(18x) ].  
\end{eqnarray}
These are the formulas that were first determined by numerical analysis \cite{Huang:2014}, and later derived analytically this way in \cite{Grassi:2014}.  Now we can easily check that our much simpler new formula for non-perturbative phase volume $\textrm{vol}_{\rm np}(E,\hbar)$ in (\ref{vol}) is equal to the GHM formula (\ref{GHMvol}) for the first few order coefficients (\ref{coefficients}). We can check the equality in the convenient variables $x$ and $Q$ here, and shall not need the formula for quantum A-period (\ref{deformedA}). The checks are performed successfully up to order $Q^7$ with our available data of Gopakumar-Vafa invariants. We conjecture that they are indeed equivalent to all orders.  This is a further support of our conjecture besides the numerical calculations of energy eigenvalues, since there are some other tests of the GHM conjecture in e.g. \cite{Kashaev:2015, Marino:2015, Kashaev:2015a}.

For the special cases when $\frac{2\pi}{\hbar} =\frac{2x}{\pi}$ is an integer, the correction term $\lambda$ vanishes, the equality $\textrm{vol}_{\rm np} = \textrm{vol}_{\rm np}^{\rm GHM} $ simply follows from the singularity cancellations with perturbative contributions. On the other hand, The equality for generic values of $\hbar$ depends on the specific data of the Gopakumar-Vafa invariants, therefore it imposes non-trivial constraints on the invariants. However, these constrains seem not strong enough to completely fix all Gopakumar-Vafa invariants. For example, the only non-vanishing Gopakumar-Vafa invariants up to $d\leq 2$ are $n^1_{0,1}=1$ and $n^2_{0,\frac{5}{2}}=1$. Assuming the vanishings $n^1_{j_L,j_R}=0$ for $j_L>0$ or $j_R>1$ and $n^2_{j_L,j_R}=0$ for $j_L>0$ or $j_R>\frac{5}{2}$, the equality $\textrm{vol}_{\rm np} = \textrm{vol}_{\rm np}^{\rm GHM} $ requires the vanishings of all Gopakumar-Vafa invariants up to $d\leq 2$ except for $n^1_{0,1}$ and $n^2_{0,\frac{5}{2}}$, though the precise non-vanishing values of $n^1_{0,1}$ and $n^2_{0,\frac{5}{2}}$ are not determined this way. In another test, we assume the vanishings of the Gopakumar-Vafa invariants as in the available data and try to constrain the non-vanishing ones with the equality $\textrm{vol}_{\rm np} = \textrm{vol}_{\rm np}^{\rm GHM} $. Up to $d\leq 7$, there are 273 non-vanishing invariants, and we find that the equality imposes 63 independent constrains. Some non-vanishing values, for example $n^3_{0,3}=1$ and $n^3_{\frac{1}{2},\frac{9}{2}}=1$, can be completely fixed by the constrains.

Our proposal can be straightforwardly generalized to other well-known toric Calabi-Yau geometries. Here we consider the cases of local Calabi-Yau threefolds which are line bundles over Hirzebruch surfaces $\mathbb{F}_0 = \mathbb{P}^1\times \mathbb{P}^1, \mathbb{F}_1, \mathbb{F}_2$. The refined Gopakumar-Vafa invariants $n^{d_1,d_2}_{j_L,j_R}$ are labelled by two degrees $d_1, d_2$, which are the degrees of  the base and fiber $\mathbb{P}^1$'s of the Hirzebruch surfaces respectively.  It is known \cite{Huang:2013} that the $ \mathbb{F}_0, \mathbb{F}_2$ models are related geometrically
\begin{eqnarray} \label{F0F2}
(n^{d_1,d_2}_{j_L,j_R})_{ \mathbb{F}_2} = \left\{
\begin{array}{cl}
0,     &   ~d_2<d_1;   \\
(n^{d_1,d_2-d_1}_{j_L,j_R})_{ \mathbb{F}_0},              &  ~d_2\geq d_1.    
\end{array}
\right.
 \end{eqnarray}

Since the mirror curve is elliptic, there is only one dynamical combination of the K\"ahler parameters  \cite{Huang:2013}, which corresponds to the quantum mechanical Hamiltonian here. The other non-dynamical combination can be treated a mass parameter. The corresponding Hamiltonians can be written as 
\begin{eqnarray}  \label{Fncurves}
\mathbb{F}_0 ~ \textrm{model}: &&  ~e^{\hat{H}} = e^{\hat{x}} + e^{\hat{p}} +  e^{-\hat{x}+m} + e^{-\hat{p}} , \\
\mathbb{F}_1 ~ \textrm{model}: &&  ~e^{\hat{H}} = e^{\hat{x}} + e^{\hat{p}} +  e^{-\hat{x}+m} + e^{-\hat{x}-\hat{p}},  \nonumber \\
\mathbb{F}_2 ~ \textrm{model}: &&  ~e^{\hat{H}} = e^{\hat{x}} + e^{\hat{p}} +  m e^{-\hat{x}} +  e^{-2\hat{x}-\hat{p}} .    \nonumber
\end{eqnarray}   
Here we require the Hamiltonian to be Hermitian and bound below in the real $(x,p)$ plane. So the mass parameter $m$ is real, and $m>-2$ for the $\mathbb{F}_2$ model.  In the followings we denote the vector degrees ${\bf d} =(d_1,d_2)$. It is also convenient to introduce a vector of integers ${\bf c} =(c_1,c_2)$ and a vector of quantum A-periods ${\bf \tilde{t}} =(\tilde{t}_1,\tilde{t}_2)$, such that the formula (\ref{fNS}) can be generalized by replacing the single degree with the dot product  
\begin{eqnarray} \label{New}
f_{NS} ({\bf \tilde{t}} , \hbar) &=& - \frac{r}{2} \sum_{j_L,j_R} \sum_{w, d_1, d_2=1}^{\infty}   (-1)^{2j_L+2j_R+rw{\bf c}\cdot{\bf d}}   n^{{\bf d} }_{j_L,j_R}      \nonumber \\ & \times & \frac{{\bf c}\cdot {\bf d}}{w} \cdot
 e^{w{\bf d}\cdot {\bf \tilde{t}}}\frac{\sin \frac{w\hbar(2j_R+1)}{2} \sin \frac{w\hbar(2j_L+1)}{2}}{ \sin^3 (\frac{w\hbar}{2})}. 
\end{eqnarray}

Now we can write down the various parameters and the perturbative phase volumes for these models.  
The results for the $F_0$ model are
\begin{eqnarray}
&&r =2,~~ {\bf c}= (1,1), ~~ {\bf \tilde{t}} = {\bf c} \tilde{t}  + (0,m), \\ \nonumber
&&\tilde{t}=-2E+(2+2e^{m})e^{-2E}+[3+8e^{m}+3e^{2m}\\ \nonumber
&& ~ +2e^{m}(e^{i\hbar}+e^{-i\hbar})]e^{-4E}+\mathcal{O}(e^{-6E}),  \\ \nonumber
&&  \textrm{vol}_{\rm p}(E,\hbar ) = \tilde{t}^2+m \tilde{t}  -\frac{2\pi^2}{3}-\frac{\hbar^2}{6} + \hbar   f_{NS} ({\bf \tilde{t}} , \hbar) .
\end{eqnarray}
The results for the $F_1$ model are
\begin{eqnarray} \label{phaseF1}
&&r =1,~~ {\bf c}= (1,2), ~~ {\bf \tilde{t}} = {\bf c} \tilde{t}  + (-m,m), \\ \nonumber
&&\tilde{t}=-E+e^{m}e^{-2E}+(e^{i\hbar/2}+e^{-i\hbar/2})e^{-3E}\\ \nonumber
&&~ +\frac{3}{2}e^{2m}e^{-4E}+\mathcal{O}(e^{-5E}),  \\ \nonumber
&&  \textrm{vol}_{\rm p}(E,\hbar ) = 4 \tilde{t}^2+m \tilde{t} -\frac{m^2}{2} -\frac{2\pi^2}{3}-\frac{\hbar^2}{6} + \hbar   f_{NS} ({\bf \tilde{t}} , \hbar)
\end{eqnarray}
The results for the $F_2$ model are
\begin{eqnarray} \label{phaseF2}
&&r =2,~~ {\bf c}= (0,1), ~~ {\bf \tilde{t}} = {\bf c} \tilde{t}  + (-2\tilde{m},\tilde{m}), \\ \nonumber
&&\tilde{t}=-2E+2 m e^{-2E}+[2+3 m^2+2(e^{i\hbar}+e^{-i\hbar})]e^{-4E}\\ \nonumber
&&
~ +\mathcal{O}(e^{-6E}),  \\ \nonumber
&&  \textrm{vol}_{\rm p}(E,\hbar ) = \tilde{t}^2-  \tilde{m}^2 -\frac{2\pi^2}{3}-\frac{\hbar^2}{6} + \hbar   f_{NS} ({\bf \tilde{t}} , \hbar) ,
\end{eqnarray}
where we introduce an additional parameter $\tilde{m} $ with the relation $2\cosh(\tilde{m}) =m$.

The various constants in the perturbative phase volumes and the formulas for scalar $\tilde{t}(E,\hbar)$  have also been determined in our previous paper \cite{Huang:2014} for the  $\mathbb{F}_0, \mathbb{F}_1$ models with the special mass $m=0$, and here we present the more general results which are obtained by the same method.  Now we can simply read off the non-perturbative phase volumes by the transformations of parameters ${\bf \tilde{t}}\rightarrow \frac{2\pi {\bf \tilde{t}}}{\hbar}, \hbar\rightarrow \frac{4\pi^2}{\hbar}$ as in (\ref{vol}) in the instanton contribution $f_{NS}$. We check the quantization condition (\ref{BScondition}) with the corresponding quantum phase volumes by some high precision numerical analysis. 

The Hamiltonian for the $\mathbb{F}_1$ model reduces to the local $\mathbb{P}^2$ model (\ref{localP2}) in the limit $m\rightarrow -\infty$. The refined Gopakumar-Vafa invariants are also related $(n^{d,d}_{j_L,j_R})_{ \mathbb{F}_1} =(n^{d}_{j_L,j_R})_{ \mathbb{P}^2} $ \cite{IKV}. In this limit the classical term in (\ref{phaseF1}) blows up, and the instanton contributions also go to infinity due to the only non-vanishing invariant $n^{1,0}_{(0,0)}=1$ with $d_1>d_2$ of the $\mathbb{F}_1$ model. There seems no simple regularization that gives a finite result, which would agree with the formula (\ref{vol}) of the  local $\mathbb{P}^2$ model. So the local $\mathbb{P}^2$ model should be treated separately.

We can also observe the correspondence between the $\mathbb{F}_0$ and $\mathbb{F}_2$ models in the case of $m_{\mathbb{F}_2}>2$, which has been pointed out in the recent papers \cite{Kashaev:2015a, Gu:2015}. 
Consider the following transformations from the $\mathbb{F}_2$ to $\mathbb{F}_0$ model
\begin{eqnarray}
&& \tilde{m} \to  \frac{1}{2}m_{\mathbb{F}_0},  ~ {\rm i.e.} ~m_{\mathbb{F}_2} \to 2\cosh(\frac{m_{\mathbb{F}_0}}{2})  ,
\nonumber \\ &&
\hat{x} \to  -\frac{1}{4}m_{\mathbb{F}_0}+\log (e^{\hat{x}}+e^{\hat{p}}), 
\nonumber \\ &&
\hat{p} \to  \hat{p}-\hat{x}+\frac{3}{4}m_{\mathbb{F}_0} -\log (e^{\hat{x}}+e^{\hat{p}}), 
\nonumber \\ &&
\hat{H} \to  \hat{H} -\frac{1}{4}m_{\mathbb{F}_0}. 
\end{eqnarray} 
Using the Baker-Campbell-Hausdorff formula, one can show that the quantum commutation $[\hat{x},\hat{p}]=i\hbar$ is preserved under the transformation, and the $\mathbb{F}_2$ quantum curve is mapped to the $\mathbb{F}_0$ quantum curve in (\ref{Fncurves}).  So the quantum spectra of the $\mathbb{F}_2$ and $\mathbb{F}_0$ models are simply related by a shift. Meanwhile, the shift in energy $E\to E -\frac{1}{4}m_{\mathbb{F}_0}$ and the transformation of the mass parameter results in a shift of the quantum A-period $\tilde{t}\to \tilde{t} + \frac{1}{2}m_{\mathbb{F}_0}$. Using the relations of Gopakumar-Vafa invariants (\ref{F0F2}), we can easily check that our proposed quantum phase volume of the  $\mathbb{F}_2$ model is then mapped to that of the  $\mathbb{F}_0$ model. 

In the case of $-2<m_{\mathbb{F}_2}<2$, the parameter $\tilde{m}$ is purely imaginary. We check that the phase volume (\ref{phaseF2}) is real and still valid. In fact, the quantization condition for ABJM model studied in \cite{Kallen:2013} corresponds to a mass parameter in this range.

Finally, we would like to address how to generalize the GHM conjecture to compare with our much simplified formula.  The formulas in (\ref{formula7}) and (\ref{GHMvol}) are generalized by the replacements $d\to {\bf c}\cdot {\bf d}$, $Q^{wd}\to \exp (\frac{2\pi}{\hbar} w {\bf d}\cdot {\bf \tilde{t}})$ and the relevant sums are now over all degrees $d_1, d_2$. The equation (\ref{formula9}) becomes  
\bes
&& \sum_{n=0}^{\infty}  \sin [\frac{2C}{3} n(n + 1)(2n + 1)x + f_s(n) + (n + \frac{1}{2})\lambda ]  \nonumber \\ &\times & Q^{\frac{Cn(n + 1)}{r}}e^{a(m)\frac{\pi}{\hbar} n(n+1)} (-1)^n e^{ f_c(n)} =0,  
\ens
where the constants $C$ and $a(m)$ are related to the quadratic and linear terms of the classical phase volume, and  are $C=\frac{9}{2}, 4,4,4$ and $a(m)=0, 2m, m, 0$ for the $\mathbb{P}^2,\mathbb{F}_0, \mathbb{F}_1, \mathbb{F}_2$ models respectively. Again, the series expansions show that the $\lambda$ corrections in the GHM quantization condition agree exactly with our new conjecture, and the equivalence imposes non-trivial constrains on the Gopakumar-Vafa invariants. 

The GHM conjecture is certainly a very remarkable tool for studying many aspects of the spectral theory. It is actually more naturally and simply written in terms of the modified grand potential, and looks more complicated here only in the form of quantization condition. Moreover, the fermionic spectral trace of the operator $\exp(-\hat{H})$ can be written as a matrix integral, whose t'Hooft expansion reproduces the conventional topological string amplitude near the conifold point \cite{Marino:2015, Kashaev:2015a}. This provides a beautiful non-perturbative formulation for the conventional, i.e. unrefined topological string theory, and is again a supporting evidence for the GHM conjecture. On the other hand, although our new formulation is much less useful in these aspects, our advantage is that we use only the Nekrasov-Shatashvili limit for both the perturbative and non-perturbative contributions. Recently, in an important progress in the studies of non-perturbative effects in conventional quantum mechanics \cite{Dunne}, it is realized that non-perturbative physics is actually often determined purely by perturbation theory, and the original quantization conditions \cite{ZinnJustin:1981} are greatly simplified in this way. Our formulation of the quantization condition confirms this appealing philosophy. Furthermore, the surprising connection between different limits of the refined topological string theory as uncovered by our formulation should be a promising direction for future explorations.

\

\noindent \textit{Acknowledgements}: We thank Xianfu Wang for useful discussions. We thank Marcos Mari\~{n}o, Mithat \"{U}nsal for comments after the first version of paper appears. MH thanks Sheldon Katz, Albrecht Klemm for collaborations on related topic. MH is supported by the ``Young Thousand People" plan by the Central Organization Department in China, and Natural Science Foundation of China.

\end{document}